\documentclass[
 reprint,
superscriptaddress,
 aps,
floatfix,
]{revtex4-2}

\usepackage[]{graphicx}
\usepackage{dcolumn}
\usepackage{bm}
\usepackage[colorlinks=true, allcolors=blue]{hyperref}

\usepackage{braket}
\usepackage{amsmath, amsthm, amssymb}
\usepackage{comment}

\begin{document}

\title{Cross-encoded quantum key distribution exploiting time-bin and polarization states with  qubit-based synchronization}

\author{Davide Scalcon}
\thanks{These authors contributed equally to this work.}
\affiliation{%
 Dipartimento di Ingegneria dell'Informazione, Universit\`a degli Studi di Padova, via Gradenigo 6B, IT-35131 Padova, Italy\\
}%

\author{Costantino Agnesi}
\thanks{These authors contributed equally to this work.}
\affiliation{%
 Dipartimento di Ingegneria dell'Informazione, Universit\`a degli Studi di Padova, via Gradenigo 6B, IT-35131 Padova, Italy\\
}%

\author{Marco Avesani}
\affiliation{%
 Dipartimento di Ingegneria dell'Informazione, Universit\`a degli Studi di Padova, via Gradenigo 6B, IT-35131 Padova, Italy\\
}%

\author{Luca Calderaro}
\affiliation{%
 Dipartimento di Ingegneria dell'Informazione, Universit\`a degli Studi di Padova, via Gradenigo 6B, IT-35131 Padova, Italy\\
}%
\affiliation{%
  ThinkQuantum S.r.l., 
  Via della Tecnica, 85, 
IT-36030 Sarcedo (VI), Italy\\
}%

\author{Giulio Foletto}
\affiliation{%
 Dipartimento di Ingegneria dell'Informazione, Universit\`a degli Studi di Padova, via Gradenigo 6B, IT-35131 Padova, Italy\\
}%

\author{Andrea Stanco}
\affiliation{%
 Dipartimento di Ingegneria dell'Informazione, Universit\`a degli Studi di Padova, via Gradenigo 6B, IT-35131 Padova, Italy\\
}%

\author{Giuseppe Vallone}
\affiliation{%
 Dipartimento di Ingegneria dell'Informazione, Universit\`a degli Studi di Padova, via Gradenigo 6B, IT-35131 Padova, Italy\\
}%
\affiliation{%
 Dipartimento di Fisica e Astronomia, Università degli Studi di Padova, via Marzolo 8, 35131 Padova, Italy\\
}%
\affiliation{%
 Padua Quantum Technologies Research Center, Università degli Studi di Padova\\
}%

\author{Paolo Villoresi}
\email{paolo.villoresi@unipd.it}
\affiliation{%
 Dipartimento di Ingegneria dell'Informazione, Universit\`a degli Studi di Padova, via Gradenigo 6B, IT-35131 Padova, Italy\\
}%
\affiliation{%
 Padua Quantum Technologies Research Center, Università degli Studi di Padova\\
}%

\date{\today}

\begin{abstract}
Robust implementation of quantum key distribution requires precise state generation and measurements, as well as a transmission that is resistant to channel disturbances. 
However, the choice of the optimal encoding scheme is not trivial and depends on external factors such as the quantum channel. 
In fact, stable and low-error encoders are available for polarization encoding, suitable for free-space channels, whereas time-bin encoding represent a good candidate for fiber-optic channels, as birefingence does not perturb this kind of states.
Here we present a cross-encoded scheme where high accuracy quantum states are prepared through a self-compensating, calibration-free polarization modulator and transmitted using a polarization-to-time-bin converter. A hybrid receiver performs both time-of-arrival and polarization measurements to decode the quantum states and successfully leaded to a transmission over 50 km fiber spool without disturbances.
Temporal synchronization between the two parties is performed with a qubit-based method that does not require additional hardware to share a clock reference.
The system was tested in a 12 hour run and demonstrated good and stable performance in terms of key and quantum bit error rates.  
The flexibility of our approach represents an important step towards the development of hybrid networks with both fiber-optic and free-space links.

\end{abstract}

\maketitle

\section{\label{Section:Introduction}Introduction}
Advancements in our ability to detect and manipulate single quantum objects has led to the development of quantum technologies with disruptive potential in many different areas, including computing, sensors, simulations, cryptography, and telecommunications. 
{One of the most mature among quantum technologies} is quantum key distribution (QKD), which allows distant users to generate a shared secret key with unconditional security. 
QKD is characterized by a consolidated composable security framework~\cite{Renner2005,Scarani2008} and by rapid and continuous technical advancements~\cite{Pirandola2019rev}.
In fact, several QKD field trials are being performed to demonstrate the real-world applicability of this technology~\cite{Dynes2019, Bacco2019, Avesani:21,Chen2021} and several start-ups and university spin-offs are being created to intercept the growing market demands. 

The most commonly used QKD protocol is the first one ever introduced, \textit{i.e.}, the BB84 protocol~\cite{Bennett2014_BB84}. 
It requires a transmitter, Alice, to send qubits encoded in two mutually unbiased bases.  
Then, a receiver, Bob, chooses an orthogonal basis for each received qubit and performs projective measurements.
After correlating their results and performing classical post-processing, Alice and Bob end up with identical keys that can be securely used in cryptographic schemes such as the one-time pad. 

The effectiveness of BB84 implementations  depends on the choice of the photonic degree of freedom that encodes the qubits.
Common choices are the polarization and time-bin degrees of freedom.
Polarization is usually preferred for free-space QKD implementations~\cite{Gong2018,Ko2018,QCosone}, even being exploited for satellite-based QKD links~\cite{Liao2018}. 
There are three main factors that encourage the use of polarization encoding for free-space links. 
The first factor is that atmospheric transmission does not change the polarization state of the transmitted qubits~\cite{Bonato:06}.
This allows Alice and Bob to share a polarization reference frame that remains stable and eliminates the need of active components to compensate the unitary transformation introduced by the quantum channel. 
The second factor is that polarization encoders with long-term temporal stability and low intrinsic quantum bit error rate (QBER) can be designed and developed. In fact, the POGNAC polarization encoder, with an average of 0.05\%, has reported the lowest intrinsic QBER in scientific literature~\cite{Agnesi2020} while the iPOGNAC~\footnote{The iPOGNAC is object of the Italian Patent No. 102019000019373 filed on 21.10.2019 as well as of the \href{https://patentscope.wipo.int/search/en/detail.jsf?docId=WO2021078723}{International Patent Application no. PCT/EP2020/079471} filed on 20.10.2020.} reported a stable polarization output for over 24 hours~\cite{Avesani2020}.
The third factor is that polarization receivers can be easily constructed with inexpensive optical components such as polarization beam splitters (PBS), half-wave plates (HWP) and quarter-wave plates (QWP) that guarantee high extinction ratios and stable performances over time.

Unfortunately, polarization encoding has some drawbacks when propagating through a fiber channel.
This is mainly due to the random changes of the fiber birefringence introduced by ambient conditions and mechanical stress.
This causes a random rotation of the polarization and, as a consequence, increases the QBER.
In turn, it lowers the secret key rate (SKR) up to the point where no quantum secure key can be established~\cite{Ding2017}.
To prevent this, a polarization compensation system becomes essential.

To make QKD performance independent of the polarization fluctuations of the optical fiber, time-bin encoding was introduced as it exploits time-of-arrival of photons and the relative phase between time bins~\cite{Bennett1992}.
This encoding has been employed in many QKD field trials in deployed fibers~\cite{Dynes2019, Bacco2019}, as well as in the record-setting 421 km fiber QKD link demonstration of the BB84 protocol~\cite{Boaron2018}.
However, time-bin has the disadvantage of requiring phase stabilization of the interferometers which encode and decode the superposition of time bins~\cite{Makarov:04}.

In this work, we present a cross-encoded implementation of the BB84 QKD protocol where polarization is used for state encoding while time-bin in used to propagate the qubits along a quantum channel composed of 50 km long fiber spool.
The iPOGNAC polarization encoder is used to generate the states required to perform QKD, which guarantees long-term temporal stability and low intrinsic QBER.
The polarization encoding is then transformed to time-bin encoding to guarantee that the birefringence of the fiber-optic channel does not modify the quantum information.
Quantum state decoding is achieved with a hybrid QKD receiver that performs both time-of-arrival and polarization measurements. 
In addition, temporal synchronization between the transmitter and the receiver is established using the qubit-based Qubit4Sync method~\cite{Calderaro2020}, without requiring supplementary hardware with respect to what is already needed for the quantum communication. 
Our work enables the implementation of flexible QKD systems that can convert the qubit encoding to best fit the characteristics of the quantum channel and represents a step towards the development of hybrid QKD networks where both fiber and free-space links are employed.

\begin{figure*}[!ht]
\includegraphics[width=\textwidth]{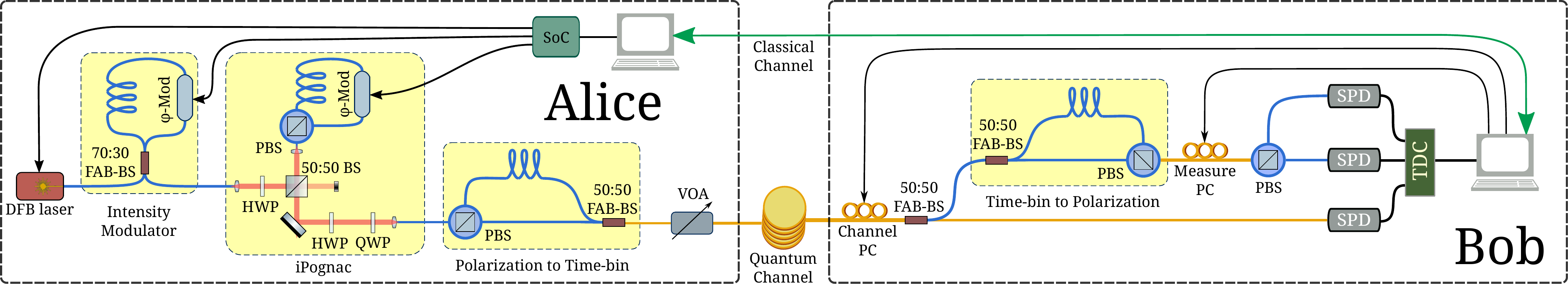}
\caption{\label{fig:setup}Experimental setup. BS: beam splitter, FAB-BS: fast-axis-blocking BS, PBS: polarization beam splitter, $\phi$-mod: phase modulator, H/QWP: half/quarter-wave plate, VOA: variable optical attenuator, PC: polarization controller, TDC: Time-to-Digital Converter, SPD: single photon detector. Single mode fibers are in yellow, polarization maintaining fibers are in blue.
}
\label{fig:Setup}
\end{figure*}

\section{\label{Section:Setup}Experimental setup}
Our cross-encoded polarization and time-bin implementation of the three-state and one-decoy efficient BB84 protocol~\cite{Grunenfelder2018} is sketched in Fig.~\ref{fig:Setup} with the transmitter, Alice, on the left and the receiver, Bob, on the right. 

\subsection{\label{Subsection:transmitter}Transmitter}
The laser source used at the transmitter is a gain-switched distributed feedback 1550 nm laser (Eblana EP1550-0-DM-H16-FM), emitting ~100 ps FWHM pulses at $R = 50 \  \rm{MHz}$ repetition rate.
The state of these light pulses is then modulated by an encoder composed of three sections: an intensity modulator, a polarization encoder and a polarization to time-bin conversion stage.
The intensity modulator is based on a fiber-optic Sagnac loop and includes a 70:30 beamsplitter (BS), a lithium-niobate phase modulator (iXBlue MPZ-LN-10), and a $1$m-long delay line~\cite{Roberts2018}.
This scheme implements the decoy state method with one decoy by setting two possible mean photon numbers (signal $\mu = 0.60$  and decoy $\nu = 0.18$) of the transmitted pulse.
These parameters are chosen in such a way that their ratio is $\mu/\nu \approx 3.33$ and the decoy intensity is sent with $P_\nu = 30\%$ probability ($P_\mu = 70\%$).

The second section, the iPOGNAC~\cite{Avesani2020}, is used to modulate the polarization state of the light.
The iPOGNAC offers fast polarization modulation with long-term stability, and a low intrinsic error rate, and, contrary to previous solutions, generates predetermined polarization states with a fixed reference frame in free-space.
Moreover, it has also been tested in a field trial in an urban environment~\cite{Avesani:21}. 
This polarization encoder relies on an unbalanced Sagnac interferometer containing a lithium-niobate phase modulator, and with the BS replaced by a fiber-based PBS with a polarization-maintaining (PM) optical fiber input and outputs.
A free-space segment (Thorlabs FiberBench), composed of a BS and a HWP, ensures the light entering the loop has the diagonal state of polarization (SOP) $\ket{D} = \left( \ket{H} + \ket{V} \right ) / \sqrt{2}$.
Hence, the light is equally split into the clockwise (CW) and counterclockwise (CCW) modes of the loop.
Thanks to the asymmetry of the interferometer, by properly setting the voltage and the timing of the pulses driving the phase modulator, one can control the SOP exiting the device as follows:
\begin{equation}
    \ket{\Phi_{\mathrm{out}}^{\phi_{\mathrm{CW}} , \phi_{\mathrm{CCW}}}} = \frac{1}{\sqrt{2}} \left ( \ket{H}  + e^{i (\phi_{\mathrm{CW}} - \phi_{\mathrm{CCW}})} \ket{V} \right )
\end{equation}
where $\phi_{\mathrm{CW}}$ and $\phi_{\mathrm{CCW}}$ are the phases applied by the phase modulator to the CW and CCW propagating light pulses.
In this experiment, the driving electric pulse amplitude was set to induce a $\pi/2$ radians phase shift, allowing the iPOGNAC to generate circular left $\ket{L} = \left( \ket{H} + i \ket{V} \right ) / \sqrt{2} $, circular right $\ket{R} = \left( \ket{H} - i \ket{V} \right ) / \sqrt{2} $ or diagonal $\ket{D}$ polarized light.
Before being coupled again into a PM optical fiber, a QWP and a HWP are used to transform circular left and right SOPs into horizontal $\ket{H}$ and vertical $\ket{V}$ SOPs.
Such transformation is achievable due to the iPOGNAC's long term stability and its ability to generate polarization states with a fixed reference frame.

Finally, the transformation of polarization encoding to time-bin is performed. This is done by a PM fiber-based unbalanced Mach-Zehnder interferometer (UMZI) where the input element is a PBS, which maps horizontal and vertical components of the light into the early and late time slots of the two dimensional time-bin encoding
\begin{equation}
    \alpha\ket{H} + \beta\ket{V} \longrightarrow \alpha\ket{E} + e^{i\phi_\mathrm{A}}\beta\ket{L}
\end{equation}
where $\phi_\mathrm{A}$ is the intrinsic phase of Alice's UMZI.
The imbalance of the MZI is approximately 2.5 ns, obtained with a 0.5 long PM fiber.
The scheme is thus able to generate the early $\ket{E}$, late $\ket{L}$ time-bin states and the superposition of the two $\ket{+} = \left (\ket{E} + e^{i\phi_\mathrm{A}} \ket{L} \right ) / \sqrt{2}$.
These states are sufficient to implement the 3-state efficient BB84 protocol~\cite{Fung2006} where the key generating basis $\mathcal{Z} = \{\ket{E} , \ket{L}\}$ is sent with 90\% probability and the control state $\ket{+}$ is sent with 10\%  probability.
The time-bin encoded signals are then attenuated down to the single-photon regime by a variable optical attenuator, then sent trough the quantum channel.

It is important to note that after the conversion to time-bin, the polarization degree-of-freedom contains no information as all the light exiting the UMZI shares the same SOP.
This is guaranteed by two factors.
First, by design, the fiber-based PBS couples the orthogonal polarization modes into the slow-axis of the PM fiber outputs.
Second, the BS used to recombine the two arms of the UMZI is a fast-axis blocking (FAB) device. FAB devices have the characteristic of discarding polarization states of the light that are aligned to the fast axis of the PM fiber, as if embedded with polarizers at both ends.

The whole system is managed by a computer, performing resource intensive tasks related to the protocol and handling classical communication. The electronic signals driving the laser and the modulators are controlled by a system-on-a-chip (SoC) which includes both a field programmable gate array (FPGA) and a CPU~\cite{Stanco2021} and is integrated on a dedicated board (Zedboard by Avnet).

\subsection{Receiver}
\begin{figure}[b!]
\includegraphics[width=0.48\textwidth]{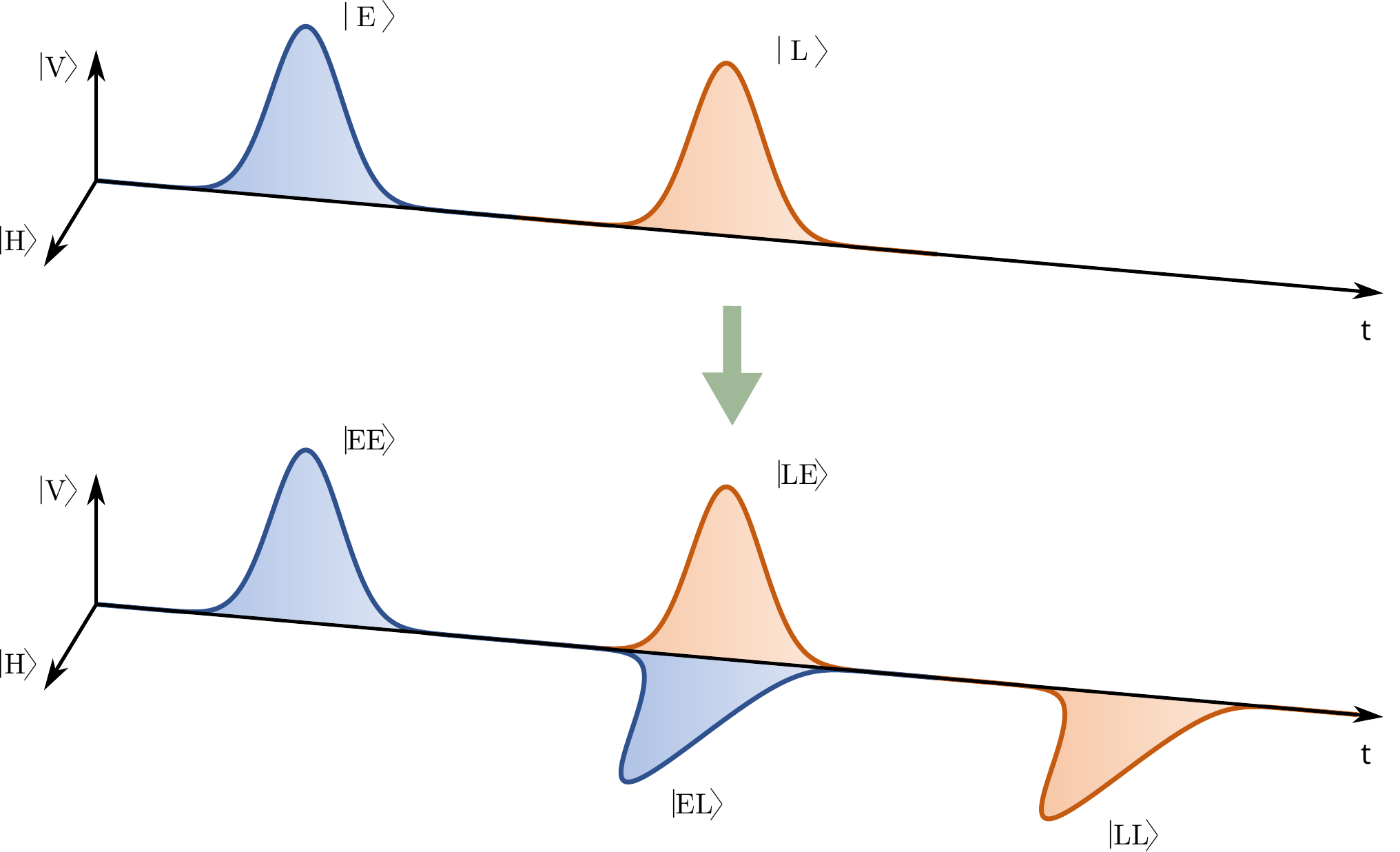}
\caption{Input and output state from the receiver's unbalanced Mach-Zehnder interferometer.
Blue and red curves represent the two possible times of emission at the transmitter.
The two lateral peaks correspond to a measurement in the key generating basis while the central peak is used to extract information on the control basis via a polarization measurement.}
\label{fig:interference}
\end{figure}

At the receiver side, the measurement basis is randomly selected by a 50:50 BS.
One of the ports is directly sent to a superconducting nanowire single photon detector (SNSPD) with approximately $80 \%$ quantum efficiency (ID281 by ID Quantique).
The overall time jtter of about $30$~ps, considering both the detector and the time-to-digital converter (quTAG by Qutools), allows the discrimination between the 2.5-ns-distant time-bins, effectively performing a measurement on the key generation basis as depicted in the upper half of Fig.~\ref{fig:interference}.
This time-of-arrival measurement has the advantage of being independent of the polarization fluctuations introduced by the fiber-optic channel, and does not require active compensation.

The other output port of the basis-selection BS is sent to an UMZI that is identical to the one used at the transmitter.
However, in this case the light is split equally between the two arms by the BS before being recombined by the PBS.
Used in this way, the UMZI outputs horizontal or vertical SOPs depending on which arm light has traveled. 
Furthermore, as depicted in the lower half of Fig.~\ref{fig:interference}, the imbalance of the UMZI temporally distributes the light in the three-peak configuration often observed in time-bin experiments.
Correspondingly, the output state from Bob's UMZI is
\begin{equation}
    \ket{\Psi_E} =  \frac{1}{\sqrt{2}} \left ( \ket{EE}\otimes\ket{V} + e^{i\phi_\mathrm{B}} \ket{EL}\otimes\ket{H} \right )
\end{equation}
when Alice transmits $\ket{E}$,
\begin{equation}
    \ket{\Psi_L} =  \frac{1}{\sqrt{2}} \left ( \ket{LE}\otimes\ket{V} +  e^{i\phi_\mathrm{B}} \ket{LL}\otimes\ket{H} \right )
\end{equation}
when Alice transmits $\ket{L}$, and
\begin{equation} \label{eq1}
\begin{split}
\ket{\Psi_+} =  \frac{1}{2} ( & \ket{EE}\otimes\ket{V} + e^{i\phi_\mathrm{B}} \ket{EL}\otimes\ket{H} \\ 
                            & + e^{i\phi_\mathrm{A}} \ket{LE}\otimes\ket{V} +  e^{i\left( \phi_\mathrm{A}+\phi_\mathrm{B}\right)} \ket{LL}\otimes\ket{H} )
\end{split}
\end{equation}
when Alice transmits $\ket{+}$, where $\phi_\mathrm{B}$ is the intrinsic phase of Bob's UMZI. 
The lateral peaks $\ket{EE}$ and $\ket{LL}$ correspond to light traveling along the short or long arms of both transmitter and receiver's UMZI and since those times-of-arrival are a measurement in the $\mathcal{Z}$ basis, they are used to generate the secret key.
Since 50\% of the light falls in these lateral peaks, by taking into account both outputs of the FAB-BS, the overall probability of measuring in the key generation basis is 75\%.
Only the central peak contains the superposition between the indistinguishable early-late $\ket{EL}$ and late-early $ \ket{LE}$ components, and the relative phase information between them is encoded in the polarization state of the light. In fact the output SOP of the central peek when $\ket{+}$ is transmitted by Alice, is 
\begin{equation}
    \ket{\psi} = \frac{1}{\sqrt{2}} \left ( \ket{H}  + e^{i \theta} \ket{V} \right ) 
\end{equation}
where $\theta = \phi_\mathrm{A} - \phi_\mathrm{B}$ is the phase difference between Alice's and Bob's UMZIs. An all-fiber electronic polarization controller (PC) composed of four piezoelectric actuators (EPC-400 by OZ Optics) is then used to transform the polarization state $\ket{\psi}$ into $\ket{D}$ state and projected in the $\{\ket{D},  \ket{A} =  \left( \ket{H} -  \ket{V} \right ) / \sqrt{2} \}$ basis. This projection is performed using a fiber PBS while the light signals are detected by two SNSPDs. Alternatively, a free-space setup with a liquid crystal, or a phase modulator with its fiber rotated by 45 degrees~\cite{Duplinskiy2017} could be used instead of the PC.
These solutions give the advantage of a simpler control scheme, due to the presence of a single degree of freedom, but would increase the losses at the receiver.

Contrary to the key generation basis, where no compensation is necessary, to perform the measurement in the control basis we need to actively compensate drifts of the relative phase shift $\theta$ of the two interferometers.
This is done by acting on the PC in front of the measurement PBS.
A coordinate descent algorithm~\cite{Wright2015} is used to minimize the measured QBER = $N_{A}/(N_D + N_{A})$ by controlling the state of the PC (labeled as Measure PC in Fig.~\ref{fig:setup}), where $N_D$ ($N_{A}$) is the number of counts in the detector associated with $\ket{D}$ ($\ket{A}$).
This algorithm, described in~\cite{Agnesi2020}, was developed for polarization tracking in polarization-encoded fiber links, and was tested in an urban QKD field trial~\cite{Avesani:21}. 
It starts operating without interrupting the QKD when the QBER exceeds 1\%, and stops when it becomes smaller than 1\%.
In our implementation the QBER is calculated rapidly by exploiting a public string of states, known to both Bob and Alice, that is interleaved with the exchange of secret qubits.
The ratio between public and secret states is 4 to 36.
However, it is important to consider that compensation in the control basis can be done without sharing any public string since the standard basis reconciliation procedure would reveal all the necessary information to estimate the QBER.
This approach would have the advantage of dedicating 100\% of time to QKD, but could be prone to some latency due to the classical communication between Alice and Bob. 

We used this hybrid time-bin to polarization scheme in the receiver to decouple the needed interferometer with the phase compensation scheme.
In fact, the phase tracking is often performed by acting on the interferometer itself, using devices like fiber stretchers~\cite{Boaron2018} or phase modulators~\cite{Wang2018} inserted in one of the optical paths.
Here, instead, the interferometer is completely passive and enclosed in a box that improves its isolation from the environment. 

A drawback of this approach is that the polarization state at the entrance of the conversion stage must be fixed and known, so that the light exits through the correct port of the closing PBS.
By manipulating the SOP in the channel, Eve could, in principle, prevent Bob from measuring the states she attacked in the control basis, thus gaining information on the key without increasing the QBER.
To avoid this, in our implementation, the basis-selection BS is FAB, meaning that only the slow-axis polarized light is measured in either basis.
In this way, Eve can no longer control the detection probability in each basis, but only the global one: if she modifies the polarization, the states do not contribute to the key and she gains no information.
This closes the security loophole but introduces some losses to the receiver, as polarization fluctuations of the input light cause variations in the detection rate.
To mitigate this effect, another PC (labeled as Channel PC in Fig.~\ref{fig:setup}) is placed in front of the receiver.
This element maximizes the total detection rate using a coordinate descent algorithm in real-time using Bob's local data without requiring any communication with the transmitter.
This PC is not involved in the measurement procedure but it is only a countermeasure to the possible degradation in the count rate due to polarization fluctuations.

The temporal synchronization is achieved using the Qubit4Sync algorithm~\cite{Calderaro2020}.
This implies that the two parties do not need a shared clock reference such as a pulsed laser~\cite{Boaron2018,Bacco2019, Dynes2019}.
Alice's clock is recovered by Bob only using the time-of-arrival of qubits while the absolute time is recovered by sending an initial public string encoded in the first $10^6$ states of the QKD transmission.
The Qubit4Sync algorithm was originally developed to work with polarization based QKD systems, making this work the first implementation of the the technique for time-bin encoded systems.

\section{\label{Section:Results}Results}

To test the performances of the developed cross-encoded QKD system, we performed a 12-hour-long QKD run exploiting a quantum channel that consisted of a 50km spool of single mode optical fiber (SM G.652.D) with 0.2 dB/km attenuation and 10 dB of additional attenuation. A summary of the main  results obtained in this experiment can be found in Table~\ref{tab:table1}. 
\begin{table}[h]
\caption{\label{tab:table1} Experimental results of the cross-encoded QKD system during the 12 hour run.}

\begin{ruledtabular}
\begin{tabular}{l|cc}
\textrm{Parameter} & \textrm{Mean value} & \textrm{Standard deviation} \\
\colrule
QBER $\mathcal{Z}$   [\%] & 0.76 & 0.08 \\
QBER $\ket{+}$ [\%]       & 0.79 & 0.65 \\
SKR [kbps]                 & 16.0 & 1.6  \\
$R_{\mathrm{det}}$ [kHz]   & 80.0 & 4.8  \\
\end{tabular}
\end{ruledtabular}

\end{table}

The mean detection rate $R_{\mathrm{det}}$ was of ~$80\cdot10^{3}$ events per seconds. Considering that on average the source emitted $(\mu P_\mu + \nu P_\nu)\cdot R  = 23.7\cdot10^{6}$ photons per second, the measured total losses were approximately 25 dB. 
The channel contribution to these losses is about 20 dB, while the remaining 5dB can be attributed to detectors efficiencies, insertion losses of optical components and fiber mating sleeves.

\begin{figure}
\includegraphics[width=0.48\textwidth]{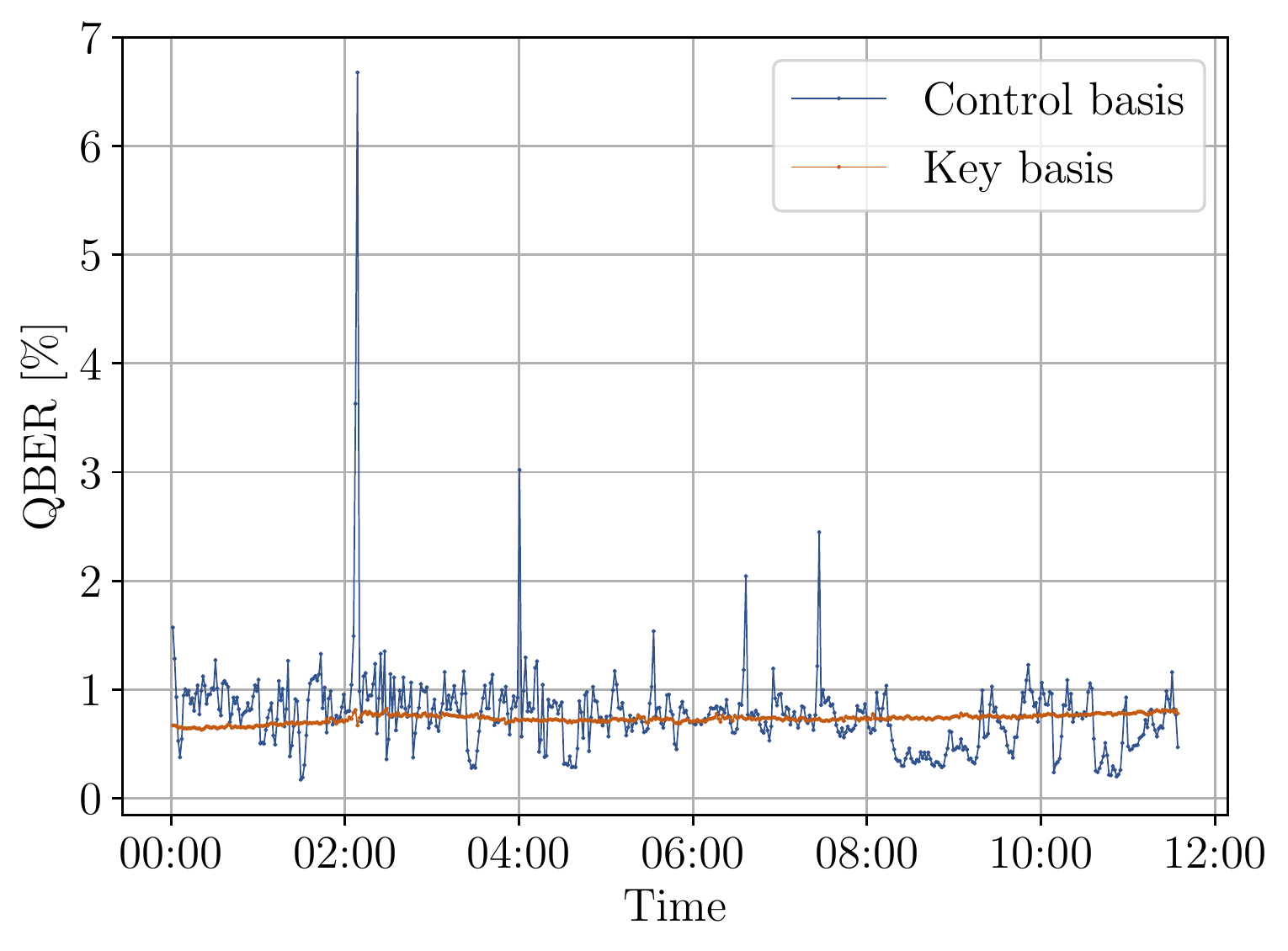}
\caption{\label{fig:QBER} Temporal evolution for the quantum bit error rate (QBER) of the key generating basis and of the control state measured every 80 seconds.
The averages are 0.765\% and 0.792\% for the key generating basis and control state respectively.}
\end{figure}

\begin{figure}
\includegraphics[width=0.48\textwidth]{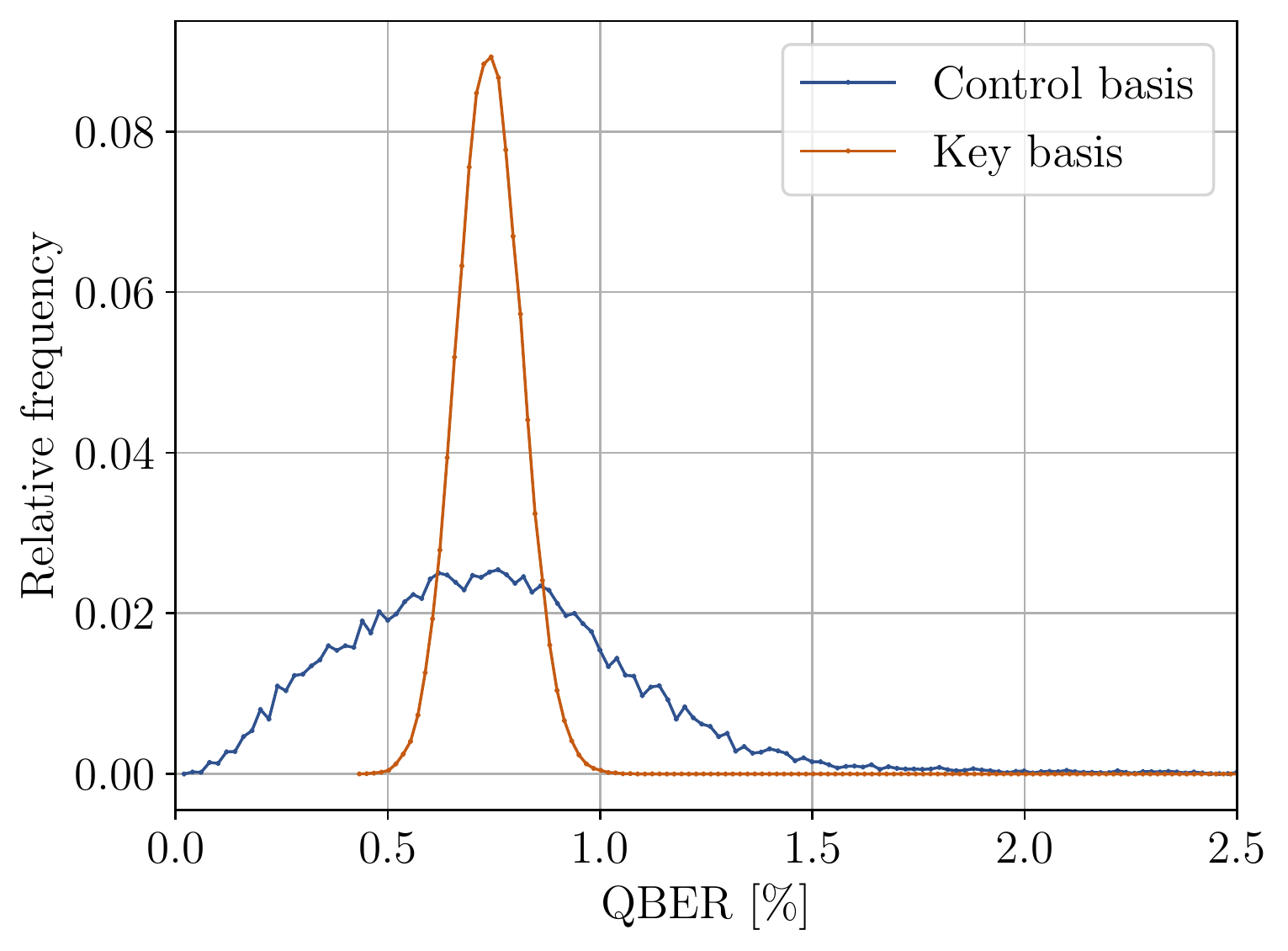}
\caption{\label{fig:QBER_DISTR} Histogram of the distribution of the quantum bit error rate (QBER) of the key generating basis and of the control state.
}
\end{figure}

The temporal evolution of the QBER on the key generation basis and on the control state is reported in Fig.~\ref{fig:QBER}, while in Fig.~\ref{fig:QBER_DISTR} their distribution is reported.
The $\mathcal{Z}$ basis QBER averages 0.765\% and remains stable throughout the whole experimental run, with a standard deviation of the 0.078\%. The control basis QBER takes greater values, with an average of 0.792\%, and distributes over a wider range, with a standard deviation of 0.651\%.
Furthermore, it can be observed that the  $\mathcal{Z}$ basis QBER is $ \leq 1\%$ for more than 99.8\% of the time without any compensation, while the control state QBER  is $ \leq 1\%$ for 81\% of the time, and $ \leq 2.5\%$ for  99.2\% of the time.
These results certify the stability of our system and its capacity of correcting the phase drifts of the UMZIs.

The $\mathcal{Z}$ basis QBER stability is inherited from the characteristics of the iPOGNAC polarization modulator used to encode the qubit states, as well as to the resistance to fluctuations of time-bin encoding.
This also demonstrates the robustness of the Qubit4Sync temporal synchronization method, which enabled highly accurate time-of-arrival measurements.  
On the other hand, fluctuations are observed for the control state QBER, mainly caused by phase drifts of the UMZIs.
However, our polarization tracking techniques effectively compensated these drifts, without ever interrupting the QKD.

\begin{figure}
\includegraphics[width=0.48\textwidth]{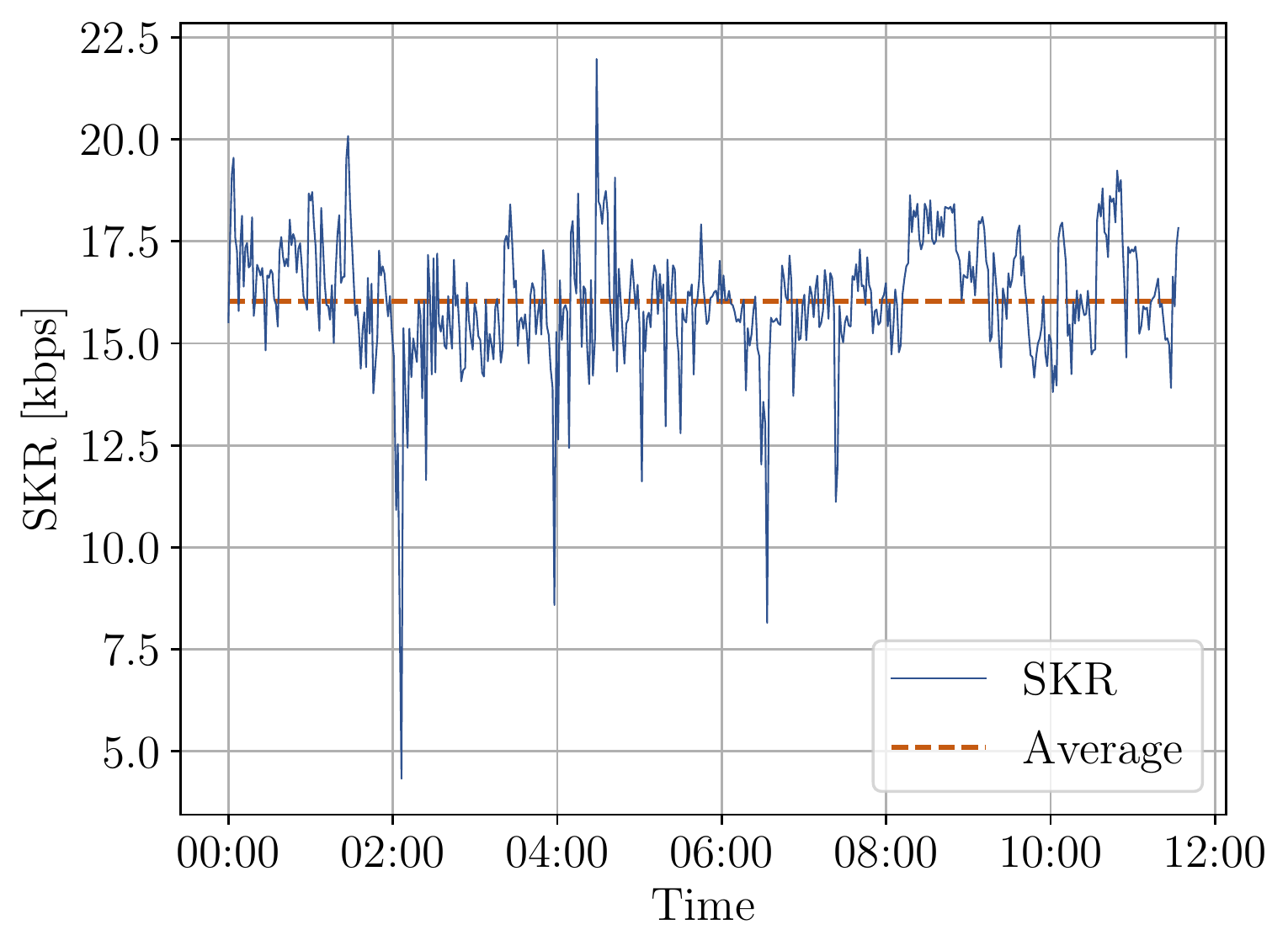}
\caption{\label{fig:SKR} Secret key rate (SKR) measured on sifted key blocks of $4\cdot 10^6$ bits (corresponding to approximately 80 seconds of acquisition). An average rate of around 16 kbps was observed.
}
\end{figure}

The post-processing uses a modified version of the AIT QKD R10 software suite~\cite{AIT} following the finite-size analysis of Ref.~\cite{Rusca2018_APL}
\begin{equation}
 \mathrm{SKR} = \frac{1}{t}\left[s_{0} + s_{1}(1  - h(\phi_\mathcal{Z})) - \lambda_{\rm EC} -\lambda_{\rm c} - \lambda_{\rm sec}\right]
 \label{eq:skr}
\end{equation}
where terms $s_{0}$ and $s_{1}$ are the lower bounds on the number of vacuum and single-photon detection events in the key generating $\mathcal{Z}$  basis, $\phi_\mathcal{Z}$ is the upper bound on the phase error rate in the $\mathcal{Z}$ basis corresponding to single photon pulses, $h(\cdot)$ is the binary entropy, $\lambda_{\rm EC}$ and $\lambda_{\rm c}$ are the number of bits published during the error correction and confirmation of correctness steps, $\lambda_{\rm sec} = 6 \log_2(\frac{19}{\epsilon_{\rm sec}})$ with $\epsilon_{\rm sec}= 10^{-10}$ is the security parameter associated to the secrecy analysis, and finally $t$ is the duration of the quantum transmission phase.
Equation \eqref{eq:skr} is applied to $4\cdot 10^6$-bit-long key blocks, a value that was chosen to produce new secret keys at a rapid pace, approximately every 80 seconds.
Increasing this value by a factor of 10 would have improved the SKR by about 20\%, at the cost of a much higher delay between the beginning of the experiment and the production of the first key.
The SKR obtained during the experiment is shown in Fig.~\ref{fig:SKR}.

\begin{figure}[tb!]
\includegraphics[width=0.48\textwidth]{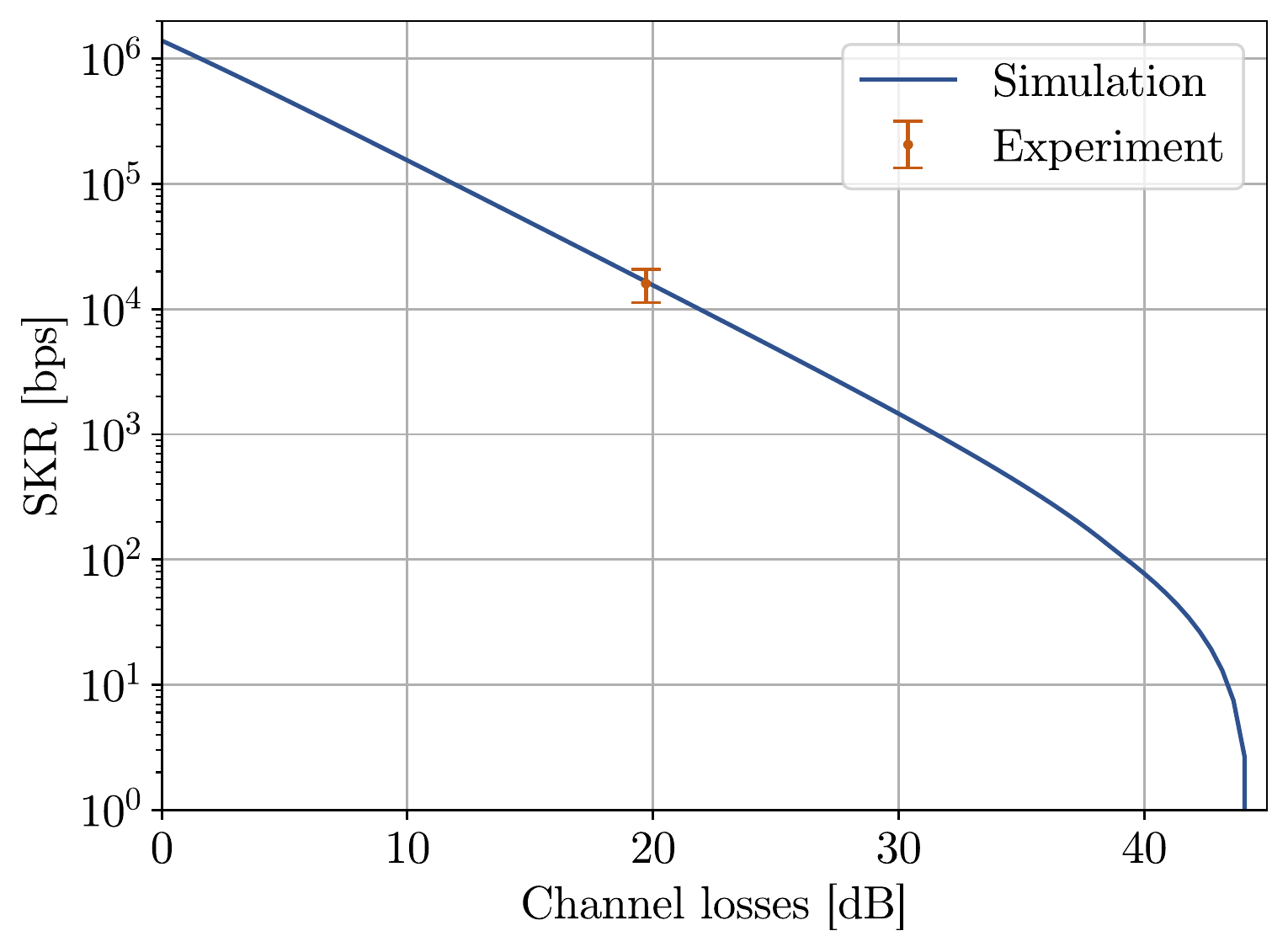}
\caption{\label{fig:Sim} Simulation of the SKR as function of the channel losses.
All other physical parameters are fixed and depend on the features of the experimental setup.
The error bar associated to the experimental data point represents three times the standard deviation.}
\end{figure}

It can be observed that our cross-encoded QKD system successfully generated secure keys without interruptions throughout the 12 hours of the experimental run and achieved an average SKR of around 16 kbps.
This result is consistent with our simulation of the performance of the system, which also predicts its behavior for different values of the channel losses, shown in Fig. \ref{fig:Sim}.
The simulation makes the strong assumption that the compensation mechanisms maintain their good performance also in conditions of strong losses, but this is in agreement with previous experiments in which the same algorithms were used for polarization correction and synchronization \cite{Agnesi2020,Calderaro2020}.

\section{\label{Section:Conclusions}Conclusions}
In this work we described a novel cross-encoded QKD scheme, based on the conversion between time-bin and polarization degrees of freedom, that implements the one-decoy, three-state BB84 protocol~\cite{Grunenfelder2018}.
By exploiting the temporally stable iPOGNAC polarization encoder we obtained polarization qubits with low error~\cite{Avesani2020}, that were converted to time-bin to allow transmission that is immune to the birefringence of the fiber-optic channel.
We implemented a hybrid receiver that performed time-of-arrival measurements for key generation as well as polarization measurements for the control states.
Temporal synchronization was successfully achieved with the Qubit4Sync method~\cite{Calderaro2020} making our work  the first implementation of time-bin encoded QKD that does not require dedicated hardware to share a temporal reference between the transmitter and the receiver. 
The developed system was tested on a 12 hours run using a 50 km fiber spool, showing a stable QBER of 0.765\% in the key basis and 0.792\% in the control state, and achieving an average SKR of of approximately 16 kbps without interruptions. 

This scheme can represent an important enabling technology for the envisioned continental-scale hybrid quantum networks that employ both fiber-optics and free-space links~\cite{Wehnereaam9288}.
In fact, since the qubit modulation of our transmitter is based on the iPOGNAC, it can be promptly reconfigured to transmit polarization-encoded qubits for free-space scenarios or, as demonstrated in this work, to convert them to time-bin for efficient propagation in an optical fiber.
In this way our transmitter is compatible with any quantum channel and the best possible encoding scheme can be chosen according to the characteristics of the link.

\begin{acknowledgments}
    
     {\noindent Author Contributions: C.A., M.A., G.V., P.V. designed the transmitter. C.A., D.S., M.A. designed the receiver. A.S., M.A., D.S. developed the transmitter electronics and the FPGA-based control system. L.C., D.S., C.A. developed the transmitter and receiver control software. G.F. developed the post-processing and simulation software. D.S. performed the experiment. All authors discussed the results. C.A., D.S. wrote the manuscript with inputs from all the authors.
     
     \noindent Part of this work was supported by: MIUR (Italian Minister for Education) under the initiative ''Departments of Excellence'' (Law 232/2016); Agenzia Spaziale Italiana (2018-14-HH.0, 
     CUP: E16J16001490001, {\it Q-SecGroundSpace}; 2020-19-HH.0,
     CUP: F92F20000000005, {\it Italian Quantum CyberSecurity I-QKD}). The AIT Austrian Institute of Technology is thanked for providing the initial elements of the post-processing software used here.}
\end{acknowledgments}

\bibliographystyle{apsrev4-2}
\bibliography{apssamp}

\end{document}